\documentclass[11pt]{article}
\usepackage{amssymb,amsmath,amsfonts,amsthm,enumerate,color}
\usepackage[toc,page,title,titletoc,header]{appendix}
\usepackage{graphicx}
\usepackage{indentfirst}
\usepackage{multicol}
\usepackage{booktabs}
\usepackage{cases}
\usepackage{tikz}

\usepackage{float}
\usepackage{subfigure}

\numberwithin{equation}{section}

\def \Vh0{\stackrel{\circ}{V}_h}

  \def\ss{\smallskip}

\newcommand{\lc}
{\mathrel{\raise2pt\hbox{${\mathop<\limits_{\raise1pt\hbox
{\mbox{$\sim$}}}}$}}}

\newcommand{\gc}
{\mathrel{\raise2pt\hbox{${\mathop>\limits_{\raise1pt\hbox{\mbox{$\sim$}}}}$}}}

\newcommand{\ec}
{\mathrel{\raise2pt\hbox{${\mathop=\limits_{\raise1pt\hbox{\mbox{$\sim$}}}}$}}}

\def\bb{\begin{equation}} \def\ee{\end{equation}}

\def\beqn{\begin{eqnarray}}  \def\eqn{\end{eqnarray}}

\def\beqnx{\begin{eqnarray*}} \def\eqnx{\end{eqnarray*}}

\def\bn{\begin{enumerate}} \def\en{\end{enumerate}}

\def\bd{\begin{description}} \def\ed{\end{description}}

\makeatletter
\newenvironment{tablehere}
  {\def\@captype{table}}
  {}
\newenvironment{figurehere}
  {\def\@captype{figure}}
  {}
\makeatother

\begin{document}

\author{
Yu Chen\thanks{School of Mathematics, Shanghai University of Finance and Economics, 777 Guoding Road, Shanghai 200433, P.R. China.  ({\tt yuchen@sufe.edu.cn}).}\,,\quad
Jin Cheng\thanks{School of Mathematical Sciences, Fudan University, Shanghai, 200433, China, P. R. China. ({\tt jcheng@fudan.edu.cn}).}\,,\quad
Yu Jiang\thanks{School of Mathematics, Shanghai University of Finance and Economics, 777 Guoding Road, Shanghai 200433, P.R. China.  ({\tt jiang.yu@mail.shufe.edu.cn}).}\quad and \;
Keji Liu\thanks{School of Mathematics, Shanghai Key Laboratory of Financial Information Technology, Shanghai University of Finance and Economics, 777 Guoding Road, Shanghai 200433, P.R. China.
 ({\tt liu.keji@sufe.edu.cn; kjliu.ip@gmail.com}).}
}

\title{\bf A Time Delay Dynamic System with External Source for the Local Outbreak of 2019-nCoV}
\date{}
\maketitle\textbf{Abstract.}
How to model the 2019 CoronaVirus (2019-nCov) spread in China is one of the most urgent and interesting problems in applied mathematics. In this paper, we propose a novel time delay dynamic system with external source to describe the trend of local outbreak for the 2019-nCoV. The external source is introduced in the newly proposed dynamic system, which can be considered as the suspected people travel to different areas. The numerical simulations exhibit the dynamic system with the external source is more reliable than the one without it, and the rate of isolation is extremely important for controlling the increase of cumulative confirmed people of 2019-nCoV. Based on our numerical simulation results with the public data, we suggest that the local government should have some more strict measures to maintain the rate of isolation. Otherwise the local cumulative confirmed people of 2019-nCoV might be out of control.

\ss

\ss

\section{ Introduction}\label{sec:intro}
In late December 2019, a cluster of serious pneumonia cases in Wuhan was caused by a novel coronavirus, and the outbreak of pneumonia began to attract considerable attention in the world.
Coronaviruses are enveloped nonsegmented positive-sense RNA viruses belonging to the family Coronaviridae and the order Nidovirales which are discovered and characterized in 1965 and are broadly distributed in humans and other mammals. In humans, most of the coronaviruses cause mild respiratory infections, but rarer forms such as the ``Severe Acute Respiratory Syndrome'' (SARS) outbreak in 2003 in China and the ``Middle East Respiratory Syndrome'' (MERS) outbreak in 2012 in Saudi Arabia and outbreak in 2015 in South Korea had cased more than 10000 cumulative cases.
In more details, there are more than 8000 confirmed SARS cases and 2200 confirmed MERS 2000 cases separately.
Although a lot of coronaviruses had been identified and characterized, they might be a tip of the iceberg and lots of potential severe and novel zoonotic coronaviruses needed to be revealed.

The World Heath Organization (WHO) designated the causative agent as the 2019 novel coronavirus (2019-nCoV), which was identified by the Chinese authorities.
Because Wuhan is the capital of Hubei province and the 7th largest city China and the largest transport hub in the central part of China, it transports millions of people to lots of cities in China and many countries in the world everyday.
Based on the special location and transport hub of Wuhan, the Chinese government has revised the law provisions of infectious diseases to add the novel 2019-nCoV as class {\bf A} agent on January 20th 2020. Moreover, a series of non-pharmaceutical interventions were implemented, say, rigorous isolation of symptomatic, suspected person, 14 days of isolation for the people who traveled from one city to another, strictly prohibit the travel in the many provinces (especially the Hubei province), the public transport is partially shut down in lots of cities, etc. However, the effectiveness and efficiency of these interventions during the early stage is questionable.
So far, there are more than 8000 confirmed cases in Wuhan and more than 24000 confirmed cases in China, and the cumulative confirmed cases  of 2019-nCoV from January 23rd to February 4th in Wuhan and mainland China are shown in Figure \ref{fig:demo}(a) and Figure \ref{fig:demo}(b) respectively. In addition, several exported cases have been confirmed in many other countries including Japan, South Korea, Singapore, USA, Canada, Germany, France, UK, Spain, etc.

\begin{figurehere}

 \hfill{}\includegraphics[clip,width=0.47\textwidth]{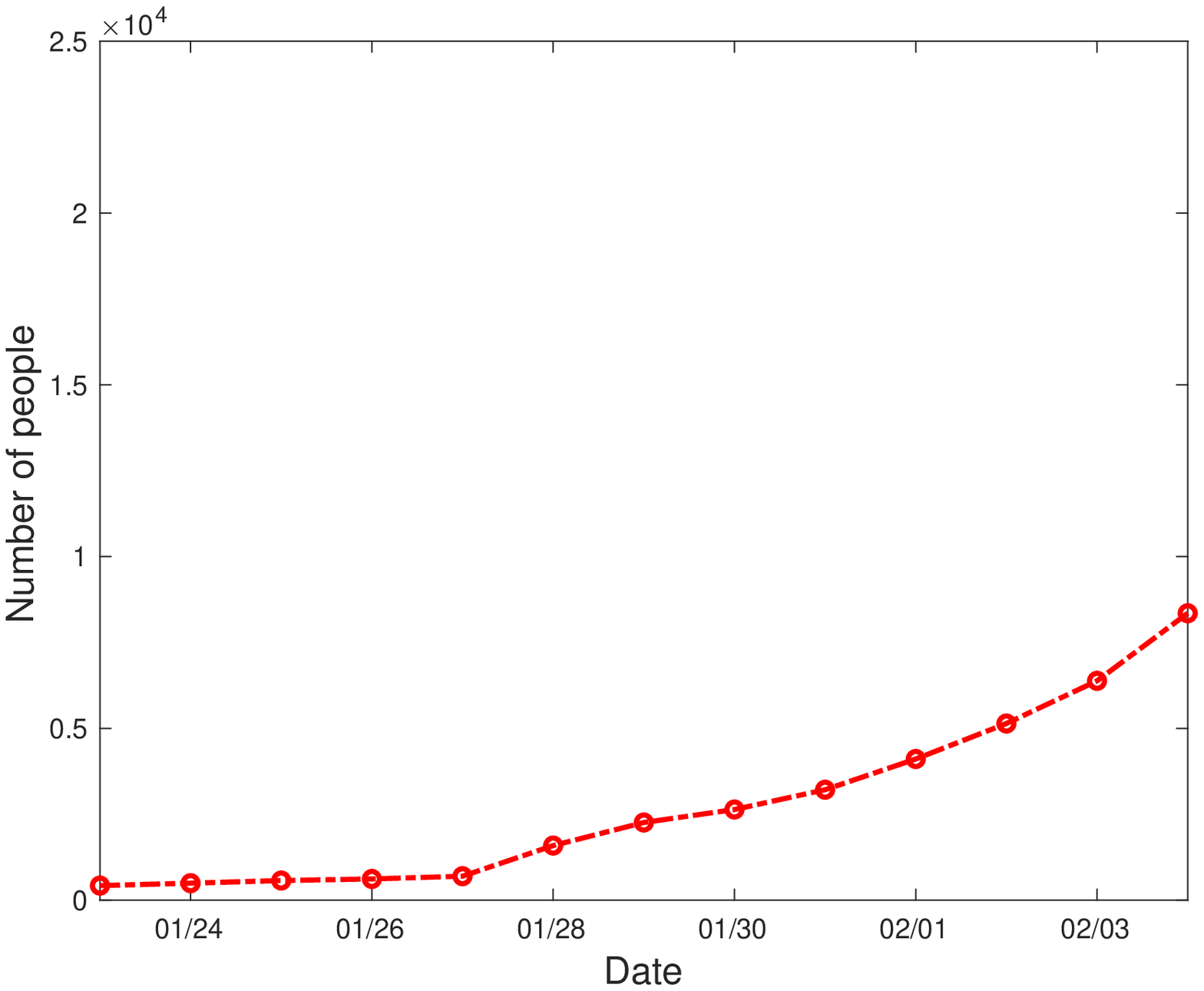}\hfill{}
  \hfill{}\includegraphics[clip,width=0.47\textwidth]{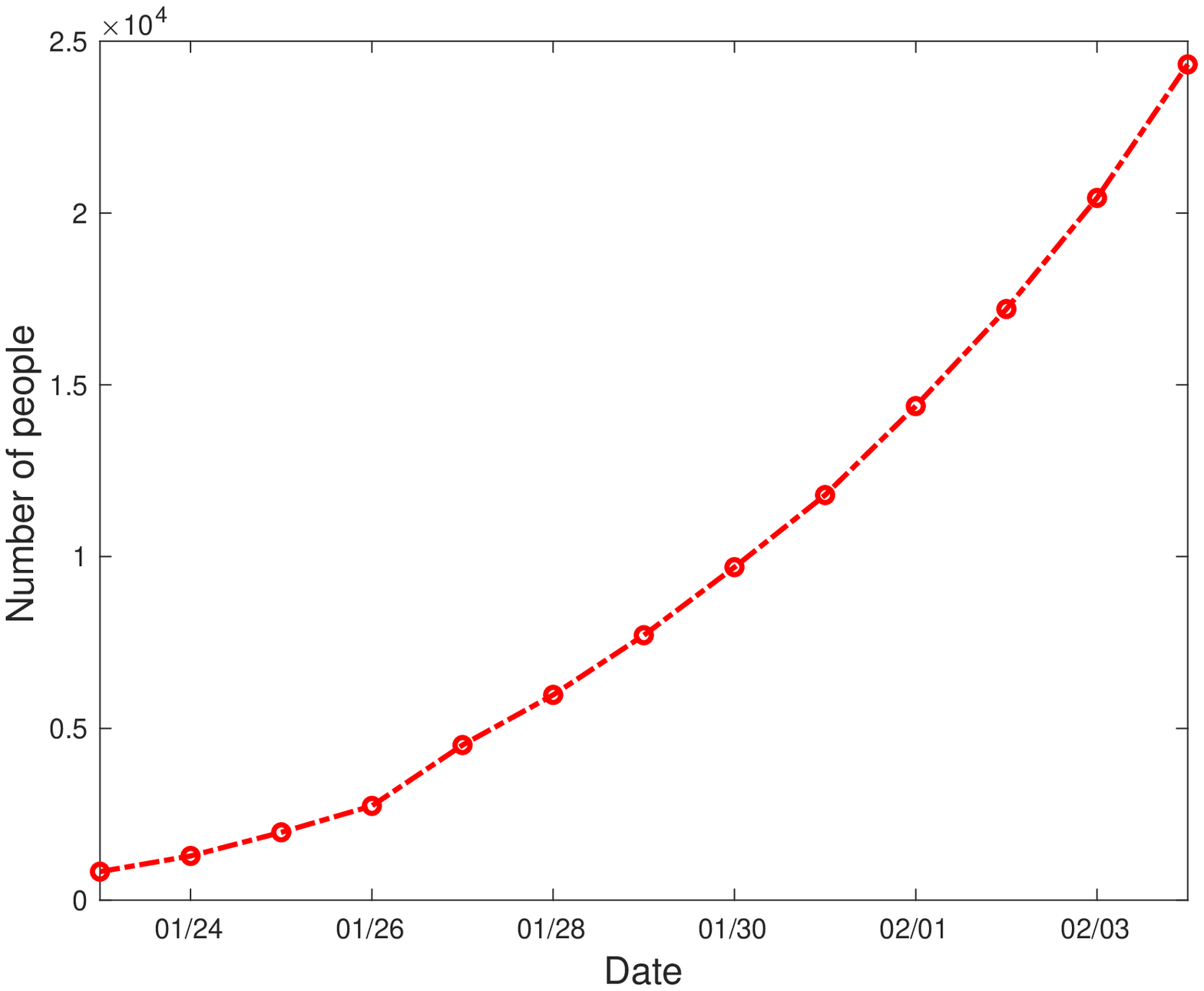}\hfill{}

 \hfill{}(a)\hfill{} \hfill{}(b)\hfill{}

 \caption{\label{fig:demo}  \small{\emph{The cumulative confirmed cases of 2019-nCoV from Jan. 23rd to Feb. 4th in Wuhan (a) and mainland China (b) respectively.}}}
 \vskip 0.3truecm
 \end{figurehere}

The WHO has recognized that the mathematical models of epidemic play a significant role in informing evidence-based decisions by health decision and policy makers.
In order to determine the impact of prevention and control of infection in different positions (i.e. provinces and cities), the strength and duration of isolation, the value for the rate of recovery,  we propose a novel dynamic system with time delay and external source in this paper.
With the help of this novel system, we are able to not only predict the trend for the outbreak of 2019-nCoV in different districts of China but also provide some helpful suggestions to achieve the maximal
protection of population with the minimal interruption of social-economic activities.

Recently we propose a novel time delay dynamical system to describe the 2019-nCoV spread in China \cite{ChenArxiv2020, Yan2020}, but it is not suitable to describe the trend of local outbreak for the 2019-nCoV. In this paper, we propose a novel time delay dynamic system with external source. In the newly proposed system, the external source term is added which can be considered as the suspected people of Wuhan (or other cities) travel to other districts of China. Moreover, we apply different kernel function in the novel system which can describe the local outbreak of 2019-nCoV more accurately.

The rest of paper is organized as follows: in section 2, we shall propose the notations, the assumptions and the corresponding novel time delay dynamic system with external source. The effective approach for estimating the parameters of novel dynamic system and the prospective cumulative confirmed people are provided in Section 3. Based on the public data, several numerical examples are exhibited in Section 4 to verify the accuracy and effectiveness of our estimation scheme and dynamic system.
Finally, we present some concluding remarks and suggestions in section 5.

\section{The Time Delay Dynamic System with External Source}
In this section, we shall state a novel dynamic system with time delay and external source to describe the local outbreak of 2019-nCoV in China.
The people in our novel dynamic system are separated into 5 kinds: external suspected people (external source), infected people, confirmed people, isolated people and cured people. Furthermore, we apply following notations to describe them,
\begin{itemize}
	\item $I(t)$: cumulative infected people at time $t$;
	\item $J(t)$: cumulative confirmed people at time $t$;
	\item $G(t)$: currently isolated people who are infected but still in latent period at time $t$;
	\item $R(t)$: cumulative cured people at time $t$.
\end{itemize}

The assumptions of external source, spread rate $\beta$, latent period $\tau_1$, delay period $t-\tau_1$, exposed people and cured rate $\kappa$ in our novel system are presented as follows:
\begin{enumerate}

	\item The transfer of infected people is assumed to be 1-to-1. And the area with external source and the area of destination would be specified in the system.
	
	\item Suppose the infected person can transfer the coronavirus to others at a spread rate $\beta$, which is defined by the average amount of people becoming infected by this person in unit time.
	
	\item In average, the infected people experience a latent period of $\tau_1$ days before they display obvious symptoms. Moreover, we assume the infected person with palpable symptoms would seek for treatment and therefore become confirmed people.
	
	\item Some of the infected people would be exposed in the latent period $\tau_1$ until they are confirmed. The average exposed period of these people are $\tau_1-\tau_1'$ days, which means they would be confirmed in the next $\tau_1'$ days. Some other part of the infected people are isolated during latent period according to investigation of diagnosed cases.
	
	\item No matter the cumulative confirmed people $J(t)$ are isolated before diagnosed or not, they are consist of the population infected at time $t-\tau_1$ averagely.
		
	\item Suppose the individual would no longer transmit the coronavirus to others when he/she is  isolated or in the treatment. Consequently, the exposed people at time $t$ are $I(t)-J(t)-G(t)$.
	
	\item It is $\tau_2$ days in average for the confirmed people become cured with rate $\kappa$ or dead with rate $1-\kappa$.
	
\end{enumerate}

With the help of the notations and assumptions stated previously, the novel time delay dynamic system with external source to describe the local outbreak of 2019-nCoV is shown in Figure \ref{fig:demo2}.


\begin{figurehere}
\vskip 0.5 truecm
\centering
\begin{tikzpicture}[>=latex,scale=1.0]
\draw[rectangle,inner sep=2.5ex] %
(0  , 5.5) node[draw] (aI)  {}%
(2  , 3.5) node[draw] (aG)  {}%
(4.2, 5.5) node       (aJ1) {}%
(4.2, 4.5) node       (aJ)  {}%
(4.2, 3.5) node       (aJ2) {}%
(7.5, 5.5) node       (aR)  {}%
(7.4, 4.5) node       (aRD) {}%
(7.5, 3.5) node       (aD)  {};%
\draw (aI) node {$\Delta I_{a}$};%
\draw (aG) node {$\Delta G_{a}$};%
\draw (aJ) node {$\Delta J_{a}$};%
\draw (3.75,5.95) rectangle (4.65,2.95);%
\draw (aR.south) node {$\substack{\displaystyle\Delta R_{a}\\[0.2ex](  \kappa)}$};%
\draw (aD.north) node {$\substack{\displaystyle\Delta D_{a}\\[0.2ex](1-\kappa)}$};%
\draw (6.95,5.95) rectangle (8.05,2.95);%
\draw[thick,->] (aI.east)  -- node[pos=0.50,above] {$\tau_{1}$ Delay} (aJ1.west);%
\draw[thick,->] (aI.south) |- node[pos=0.75,above] {$\tau_{1}-\tau_{1}^{\,\prime}$} (aG.west);%
\draw[thick,->] (aG.east)  -- node[pos=0.50,above] {$\tau_{1}^{\,\prime}$} (aJ2.west);%
\draw[thick,->] (aJ.east)  -- node[pos=0.50,above] {$\tau_{2}$ Delay} (aRD.west);%
\draw[thick,->] (aI.north) -- (0,6.7) -- node[pos=0.50,above] {$\tau_{1}+\tau_{2}$ Delay} (7.5,6.7) -- (aR.north);%
\draw[red,dashed] (-0.7,2.7) rectangle (8.3,7.4);%
\draw[rectangle,inner sep=2.5ex] %
(0  , 0) node[draw] (bI)  {}%
(2  ,-2) node[draw] (bG)  {}%
(4.2, 0) node       (bJ1) {}%
(4.2,-1) node       (bJ)  {}%
(4.2,-2) node       (bJ2) {}%
(7.5, 0) node       (bR)  {}%
(7.4,-1) node       (bRD) {}%
(7.5,-2) node       (bD)  {};%
\draw (bI) node {$\Delta I_{b}$};%
\draw (bG) node {$\Delta G_{b}$};%
\draw (bJ) node {$\Delta J_{b}$};%
\draw (3.75,0.45) rectangle (4.65,-2.45);%
\draw (bR.south) node {$\substack{\displaystyle\Delta R_{b}\\[0.2ex](  \kappa)}$};%
\draw (bD.north) node {$\substack{\displaystyle\Delta D_{b}\\[0.2ex](1-\kappa)}$};%
\draw (6.95,0.45) rectangle (8.05,-2.45);%
\draw[thick,->] (bI.east)  -- node[pos=0.50,above] {$\tau_{1}$ Delay} (bJ1.west);%
\draw[thick,->] (bI.south) |- node[pos=0.75,above] {$\tau_{1}-\tau_{1}^{\,\prime}$} (bG.west);%
\draw[thick,->] (bG.east)  -- node[pos=0.50,above] {$\tau_{1}^{\,\prime}$} (bJ2.west);%
\draw[thick,->] (bJ.east)  -- node[pos=0.50,above] {$\tau_{2}$ Delay} (bRD.west);%
\draw[thick,->] (bI.north) -- (0,1.2) -- node[pos=0.50,above] {$\tau_{1}+\tau_{2}$ Delay} (7.5,1.2) -- (bR.north);%
\draw[red,dashed] (-0.7,-2.7) rectangle (8.3,1.9);%
\draw[ultra thick,->] (3.8,2.7) -- (3.8,1.9);%
\end{tikzpicture}\label{fig:demo2}
\caption{\em The demonstration of novel time delay dynamic system with the external source.}
\vskip 0.3 truecm
\end{figurehere}

Since the novel dynamic model would be more complicated compared with the one proposed in \cite{ChenArxiv2020}, we first provide the following general form of time delay model that is valid for various cases including those with source or sink,
\begin{equation}
\left\{
\begin{aligned}
\frac{\mathrm{d}I}{\mathrm{d}t}&=\tilde{\mathcal{I}}(t)\\[2mm]
\frac{\mathrm{d}J}{\mathrm{d}t}&=\gamma \int_0^t h_1(t-\tau_1,t')\tilde{\mathcal{I}} (t')\mathrm{d}t',\\[2mm]
\frac{\mathrm{d}G}{\mathrm{d}t}&=\tilde{\mathcal{G}}(t)-\int_0^t h_2(t-\tau_2,t')\tilde{\mathcal{G}}(t')\mathrm{d}t'  \\[2mm]
\frac{\mathrm{d}R}{\mathrm{d}t}&=\kappa\int_0^t h_3(t-\tau_1-\tau_2,t')\tilde{\mathcal{I}}(t') \mathrm{d}t'.
\end{aligned}
\right.
\label{eq:newmodel-basic}
\end{equation}

In order to provide a better understanding of the dynamic system \eqref{eq:newmodel-basic}, we present some detailed explanations as follows:
\begin{enumerate}

\item $\tilde{\mathcal{I}}$ is generally the increase rate of cumulative infected people $I(t)$ at time $t$.
The specific form of $\tilde{\mathcal{I}}$ depends on the system is closed or has outflow/inflow, and it would be illustrate in detail later.

\item Because all the cumulative confirmed people $J(t)$ come from the previously infected population, the increment of $I(t)$ at time $t'$ ($t'<t$), i.e., $\tilde{\mathcal{I}}(t')$, which means the increment of $J(t)$ depends on the history of $\tilde{\mathcal{I}}(t)$. If the average delay between infected time and confirmed time is $\tau_1$, the increase rate of $J$ can be represented as
\begin{equation}
\gamma\int_0^t h_1(t-\tau_1,t')\tilde{\mathcal{I}}(t')\mathrm{d}t',\label{1delay}
\end{equation}
where $\gamma$ is the morbidity, and $h_1(\hat{t},t')$ ($\hat{t}=t-\tau_1$) is a distribution which should be normalized as
\[\int_0^t h_1(\hat{t}, t')\mathrm{d}t'=1,\quad \hat{t}\in (0,t).\]
We are able to observe that $h_1(\hat{t},t)$ can be regarded as the probability distribution of infection time $t'$, and here we take the normal distribution
\[h_1(\hat{t},t')=c_1 e^{-c_2 (\hat{t}-t')^2}\]
with $c_1$ and $c_2$ be constants.

\item The function $\tilde{\mathcal{G}}(t)$ is the newly isolated and infected people.  The integral term in the equation for $G(t)$ means the people isolated $\tau_2$ days ago (averagely) and would be confirmed and sent for treatment, who would no longer be counted into to the instant isolated population. In addition, the kernel function in the integral has the following expression
$$h_2(\hat{t},t')=c_3 e^{-c_4 (\hat{t}-t')^2}$$
with $c_3$ and $c_4$ be constants.

\item The parameter $\kappa$ is the cured rate. The time delay term is obtained similarly as that of $J(t)$'s, and $h_3(\hat{t},t')=c_5 e^{-c_6 (\hat{t}-t')^2}$ with $c_5$ and $c_6$ be constants.

\end{enumerate}

We next derive the specific form of cases with inflow or outflow sources separately, i.e. the specific form of $\tilde{\mathcal{I}}$.
For simplicity we consider the half open cases, i.e. with single outflow or single inflow. The situation with both inflow and outflow can be treated based on \eqref{eq:newmodel-basic} similarly.
It is assumed that some of the infected people of Area {\bf a} would transfer to Area {\bf b}. We employ {\bf a} and {\bf b} to distinguish the source area and destination respectively. The dynamic system for destination {\bf b} applies the output of source area {\bf a} as input. Although the cured rate and dead rate are the same for the two dynamic systems, the isolation ratio $\ell$ and infection rate $\beta$ are different for these two dynamic systems.

\paragraph{\large Area a with single outflow to Area b}

For the case with merely outward transfer, the exposed people at time $t$ is $I_a(t)-J_a(t)-G_a(t)$, which further make $\beta_a(I_a(t)-J_a(t)-G_a(t))$ people infected with $\beta_a$ as the infection rate for this area. Meanwhile, $\nu(t)\theta(I_a(t)-J_a(t)-G_a(t))$ of them transfers to other regions, where $\theta$ is the coefficient of transport activity and $\nu(t)$ is the time-dependent distribution of exposed people that are likely to move out. Consequently, at time $t$ the net increment of the infected number in this region is
\begin{equation}
\tilde{\mathcal{I}}_a(t)=\Big(\beta_a-\nu(t)\theta\Big) \Big(I_a(t)-J_a(t)-G_a(t)\Big).
\label{eq-I1}
\end{equation}
We further assume that
\begin{equation}
\tilde{\mathcal{G}}_a(t)=\ell_a  \Big(I_a(t)-J_a(t)-G_a(t)\Big)
\label{eq-G1}
\end{equation}
which means the currently exposed people are isolated at rate $\ell_a$. By substituting \eqref{eq-I1} and \eqref{eq-G1} into \eqref{eq:newmodel-basic}, we arrive at the following expressions for the single outflow system
\begin{equation}
\left\{
\begin{aligned}
\frac{\mathrm{d}I_a}{\mathrm{d}t}&=\tilde{\mathcal{I}}_a(t),\\[2mm]
\frac{\mathrm{d}J_a}{\mathrm{d}t}&=\gamma \int_0^t h_1(t-\tau_1,t')\,\tilde{\mathcal{I}}_a(t')\mathrm{d}t',\\[2mm]
\frac{\mathrm{d}G_a}{\mathrm{d}t}&=\tilde{\mathcal{G}}_a(t)
-\int_0^t h_2(t-\tau'_1,t') \,\tilde{\mathcal{G}}_a(t') \mathrm{d}t',\\[2mm]
\frac{\mathrm{d}R_a}{\mathrm{d}t}&=\kappa\int_0^t h_3(t-\tau_1-\tau_2,t')\, \tilde{\mathcal{I}}_a(t')\mathrm{d}t'.
\end{aligned}
\right.
\label{eq:newmodel-output}
\end{equation}

\paragraph{\large Area b with single inflow from Area a}
We next concern about another system with single external source from Area {\bf a}. The instant increment of infected people caused by existed exposed people at time $t$ is also assumed as $\beta_b \Big(I_b(t)-J_b(t)-G_b(t)\Big)$, while the external transport contribute to the increment is $\tilde{\mathcal{I}}_{In}(t)$, thus the total increment at that time is
\begin{equation}
\tilde{\mathcal{I}}_b(t): =\beta_b \Big(I_b(t)-J_b(t)-G_b(t)\Big)+\tilde{\mathcal{I}}_{In}(t),
\end{equation}
where $\tilde{\mathcal{I}}_{In}(t)$ is the output of source region, namely,
\begin{equation}
\tilde{\mathcal{I}}_{In}=\nu(t)\theta\Big(I_a(t)-J_a(t)-G_a(t)\Big).\label{eq:externalsource}
\end{equation}
The rate of isolation is still depends on the existed exposed amount of people,
so we possess the following form similarly as \eqref{eq-G1}
\begin{equation}
\tilde{\mathcal{G}}_b(t)=\ell_b  \Big(I_b(t)-J_b(t)-G_b(t)\Big),
\label{eq-G1}
\end{equation}
where the currently exposed people are isolated at rate $\ell_b$.
 According to \eqref{eq:newmodel-basic}, we are able to state the novel time delay dynamic system for Area {\bf b} with the external source from Area {\bf a} as follows:
\begin{equation}
\left\{
\begin{aligned}
\frac{\mathrm{d}I_b}{\mathrm{d}t}&=\tilde{\mathcal{I}}_b(t),\\[2mm]
\frac{\mathrm{d}J_b}{\mathrm{d}t}&=\gamma \int_0^t h_1(t-\tau_1,t')\,\tilde{\mathcal{I}}_b(t')\mathrm{d}t',\\[2mm]
\frac{\mathrm{d}G_b}{\mathrm{d}t}&=\tilde{\mathcal{G}}_b(t)
-\int_0^t h_2(t-\tau'_1,t') \,\tilde{\mathcal{G}}_b(t') \mathrm{d}t',\\[2mm]
\frac{\mathrm{d}R_b}{\mathrm{d}t}&=\kappa\int_0^t h_3(t-\tau_1-\tau_2,t')\, \tilde{\mathcal{I}}_b(t') \mathrm{d}t'.
\end{aligned}
\right.
\label{eq:newmodel-input}
\end{equation}

In sum, the above 1-to-1 single direction model is consistent of two sets of equations corresponding to the source system and destination system. It can be observed that the source system \eqref{eq:newmodel-output} of Area {\bf a} is independent, which means it can be solved without any information of system \eqref{eq:newmodel-input} of Area {\bf b}.  Conversely, the system \eqref{eq:newmodel-input} of Area {\bf b} relies on the source system \eqref{eq:newmodel-output} of  Area {\bf a}, so the output of Area {\bf a} is applied as its input in the computation.

\section{The Estimation and Prediction Scheme}
In this section, we shall first state the optimization method to estimate some parameters of the dynamic system \eqref{eq:newmodel-input} from the official data, and a prediction scheme would be presented to forecast the tendency of outbreak for the 2019-nCoV.

Based on the information of parameters $\{\beta_a,\kappa,\ell_a,\gamma,\tau_1,\tau'_1,\tau_2\}$ and initial conditions $\{I_a(t_0), $ $G_a(t_0), J_a(t_0),R_a(t_0)\}$ in the novel dynamic system \eqref{eq:newmodel-output}, the cumulative cured people $R_a(T)$ and the cumulative confirmed people $J_a(T)$ at any given time $T$ are readily to attain by solving the novel dynamic system \eqref{eq:newmodel-output} numerically.
Likewise, the other novel dynamic system \eqref{eq:newmodel-input} can be solved with the extra information of parameters $\{\beta_b,\ell_b\}$, the external source $\tilde{\mathcal{I}}_{In}$ \eqref{eq:externalsource} and the initial conditions $\{G_b(t_0), J_b(t_0),R_b(t_0)\}$.
In the practical applications, we suggest to apply the Matlab$\circledR$ inner-embedded program {\bf dde23} to solve the novel dynamic systems \eqref{eq:newmodel-output} and \eqref{eq:newmodel-input}.

In addition, the following conditions for the initial time and some parameters are assumed in the practical applications:
\begin{enumerate}
	
	\item{\bf initial conditions:}  On the initial day $t_0$, we suppose $5$ people in the Area {\bf a} are infected the 2019-nCoV from unknown sources. Moreover, the confirmed, isolated and recovered people are all 0 on the initial day. All these assumptions represent $I_a(t_0)=1$, $G_a(t_0)=J_a(t_0)=R_a(t_0)=0$ and $G_b(t_0)=J_b(t_0)=R_b(t_0)=0$.
In the numerical simulation, we further assume that there are no isolation measures implemented before $T=t_0+15$.
		\item {\bf parameters:}
According to the present data, we suppose a relatively high cure rate as $\kappa=0.97$, the morbidity is relatively high with $\gamma=0.99$. The average latent period $\tau_1$ and treatment period $\tau_2$ are also regarded as known according to the official data. The average period between getting isolated and diagnosed $\tau_1'$ satisfies $0<\tau_1'<\tau_1$. The known parameter set is summarized in Table \ref{par-known}.
\begin{tablehere}
	\begin{center}
		\begin{tabular}{ccccc}
			\hline
			$\kappa $&$\gamma$ &$\tau_1$ &$\tau'_1$ & $\tau_2$  \\
			\hline
			0.97 & $0.99$ & $7$ &$5$ & $12$  \\
			\hline
		\end{tabular}
	\end{center}
	\caption{\em The settings of parameters based on the official data.}\label{par-known}
\end{tablehere}
\end{enumerate}

Accordingly, the rest parts of parameters that need to be estimated are as follows,
\[\Theta_a:=[\beta_a,\ell_a]\quad\text{and}\quad \Theta_b:=[\beta_b,\ell_b],\]
and the identifications of parameters $\Theta_a$ and $\Theta_b$ come to the following two optimization problems,
\begin{equation}\label{eq:op1}
\min_{\Theta_a}\|J_a(\Theta_a;t)-J^a_{Obs}\|_2,
\end{equation}
and
\begin{equation}\label{eq:op2}
\min_{\Theta_b}\|J_b(\Theta_b;t)-J^b_{Obs}\|_2,
\end{equation}
where $J^a_{Obs}$ and $J^b_{Obs}$ are separately the daily official data in Area {\bf a} and Area {\bf b} reported by the National Health Commission of China.

Consequently, we solve the optimization problem \eqref{eq:op1} to derive the parameter $\Theta_a$ initially. In addition, with the knowledge of parameters $\Theta_a$, we are able to predict the trend of local outbreak in Area {\bf a} and the external source \eqref{eq:externalsource} for Area {\bf b}. Furthermore, solve the optimization problem \eqref{eq:op2}, we obtain the parameter $\Theta_b$ and are able to predict the tendency of local outbreak in Area {\bf b}.\

 The whole procedure is concluded as follows:\\[2mm]
{\bf The Estimation and Prediction Scheme}:
\begin{enumerate}
\item[Step 1.]  Based on the official data $J^a_{Obs}$, we apply the Levenberg-Marquad (LM) method or the Markov chain Monte Carlo (MCMC) method \cite{KNO,KS} to solve the optimization problem \eqref{eq:op1}, and the estimated parameter $\Theta_a^*$ is obtained.

\item[Step 2.] With the reconstructed $\Theta_a^*$ , one could acquire the predictions of $\{I_a(t),J_a(t), G_a(t), R_a(t)\}$ and $\tilde{\mathcal{I}}_{In}$.

\item[Step 3.]  Based on the official data $J^b_{Obs}$, we solve the optimization problem \eqref{eq:op2} and attain the estimated parameter $\Theta_b^*$ in account of $\tilde{\mathcal{I}}_{In}$.

\item[Step 4.]   The values of $\{I_b(t),J_b(t), G_b(t), R_b(t)\}$ are obtained by solving the novel dynamic system \eqref{eq:newmodel-input} numerically.

\end{enumerate}

\section{Numerical Simulations}
In this section, we shall present some numerical experiments to verify the accuracy and efficient of the estimation and prediction scheme.
It is worth mentioning that the data employed in our novel dynamic system are acquired from the Health Commission of each province and city of China and the National Health Commission of China. Moreover, the data includes the cumulative confirmed people, and the cumulative cured people and the cumulative dead people from Jan. 23rd 2020 to Feb. 4th 2020.

In order to predict the external source, we design the time distribution of exposed people moving out as
$$\nu(t)=e^{-0.1(t-t_1)^2}+e^{-0.1(t-t_2)^2},$$
which illuminates there would be two peaks around February 8th and February 20th separately.
The reason for these two peaks is that the people would go back to work and school after the Lunar New Year on the specified two days.
By implementing the estimation and prediction scheme of section 3,  the estimations of parameters are $\beta_a=0.27$ and $\ell_a=0.482$, and $\tilde{\mathcal{I}}_{In}(t)$ with diffident $\theta$ are exhibited in Figure \ref{fig:ex-source}.

\begin{figurehere}
	\begin{center}
	\vskip -0.4 truecm
		\includegraphics[width=0.5\textwidth]{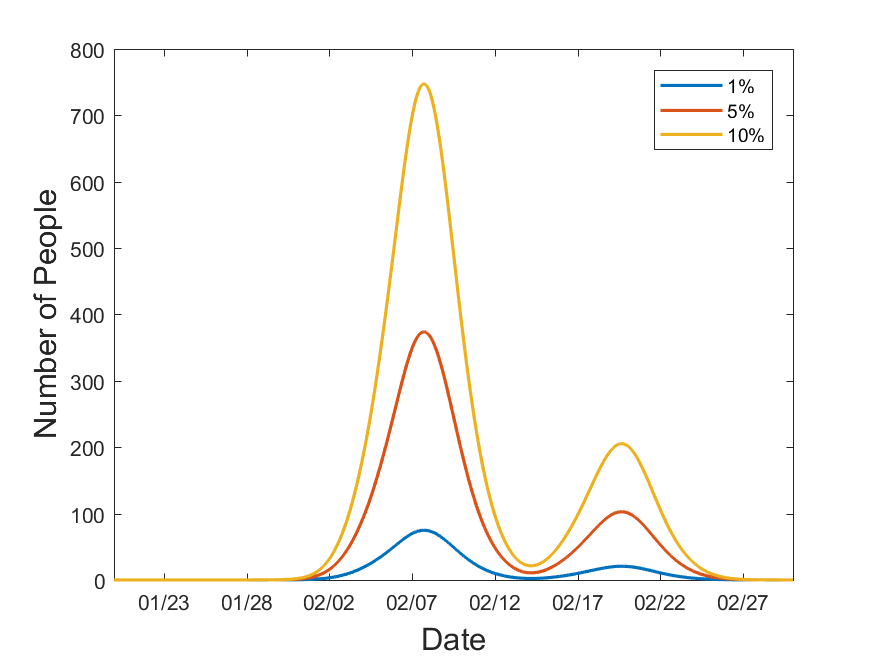}
	\end{center}
	\vskip -0.5 truecm
	\caption{\em The external source $\tilde{\mathcal{I}}_{In}(t)$ with $\theta=5\%$, 10\% and 20\%.}
	\label{fig:ex-source}
	\vskip 0.3 truecm
\end{figurehere}

For the purpose of comparing the tendency of Area {\bf b} with the external source and without it,
we first show the prediction of area {\bf b} without external source, which means the case with the parameter $\theta=0$, in Figure \ref{Ideal60}.
\begin{figurehere}
	\begin{center}
		\includegraphics[width=0.5\textwidth]{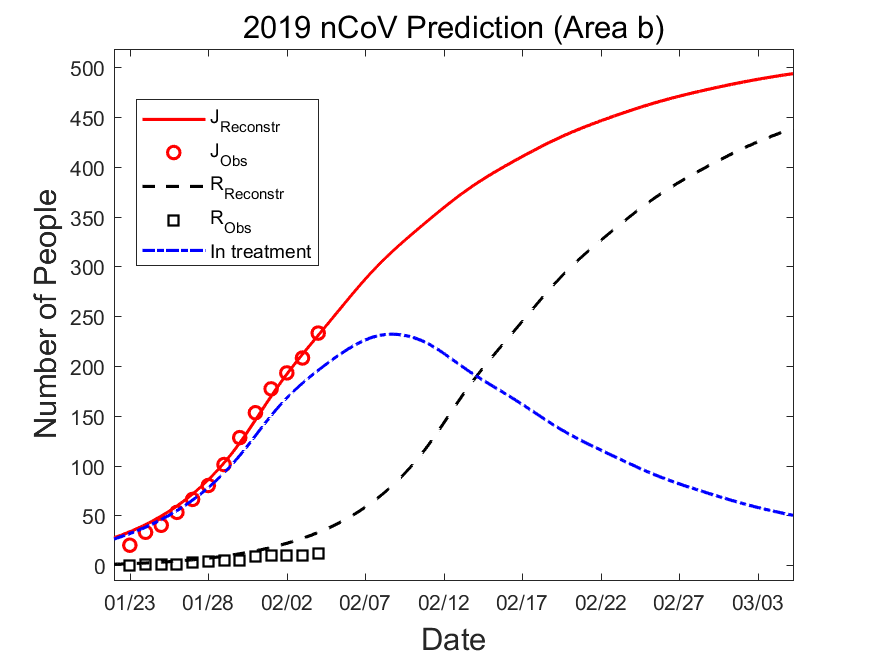}
	\end{center}
	\vskip -0.5 truecm
	\caption{\em Prediction without external source}
	\label{Ideal60}
	\vskip 0.3 truecm
\end{figurehere}
It is obviously that the situation of pneumonia would tend to end at the beginning of March and the final cumulative confirmed people would be about 500. However, by observing the prediction of Area {\bf b} with the external source in Figure \ref{Ideal5}, the increase of exposed people would lead to the increase of final cumulative confirmed people.

\begin{figurehere}
	\begin{center}
		\subfigure[$\theta=0\%$]
		{
			\includegraphics[width=0.45\textwidth]{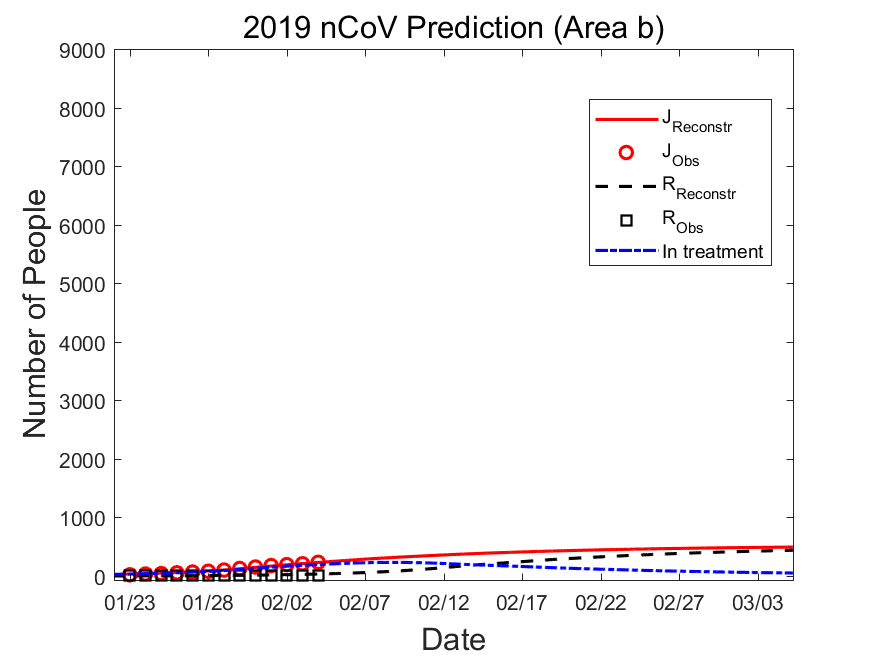}
			\label{ideal-31}
		}
		\subfigure[$\theta=1\%$]
		{
			\includegraphics[width=0.45\textwidth]{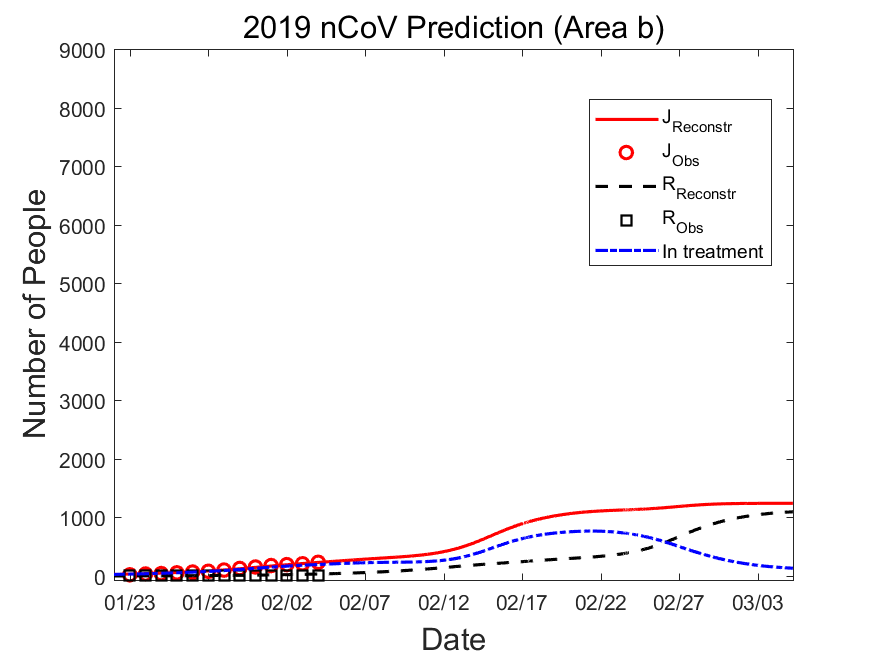}
			\label{ideal-32}
		}
		\\
		\subfigure[$\theta=5\%$]
		{
			\includegraphics[width=0.45\textwidth]{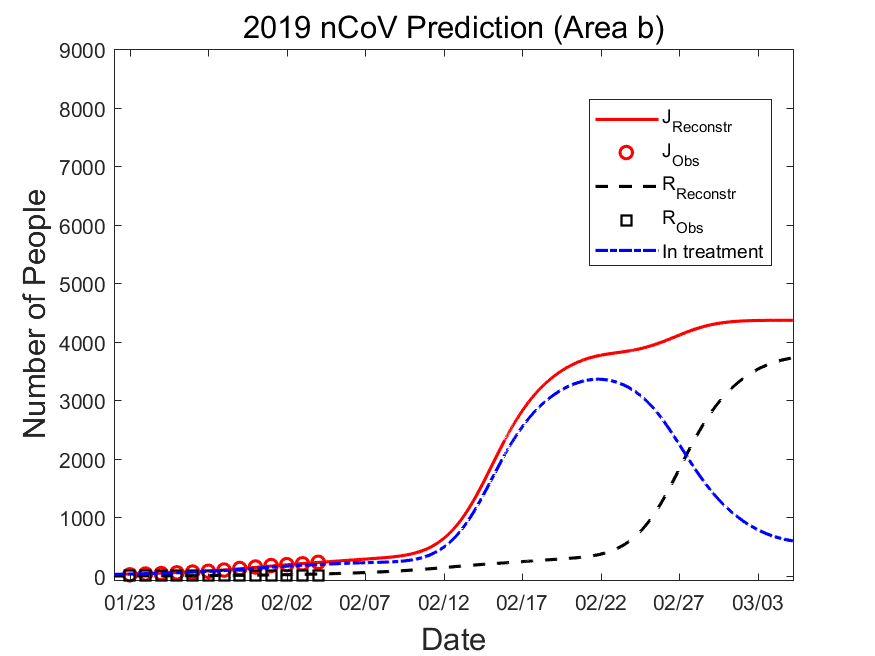}
			\label{ideal-33}
		}
		\subfigure[$\theta=10\%$]
		{
			\includegraphics[width=0.45\textwidth]{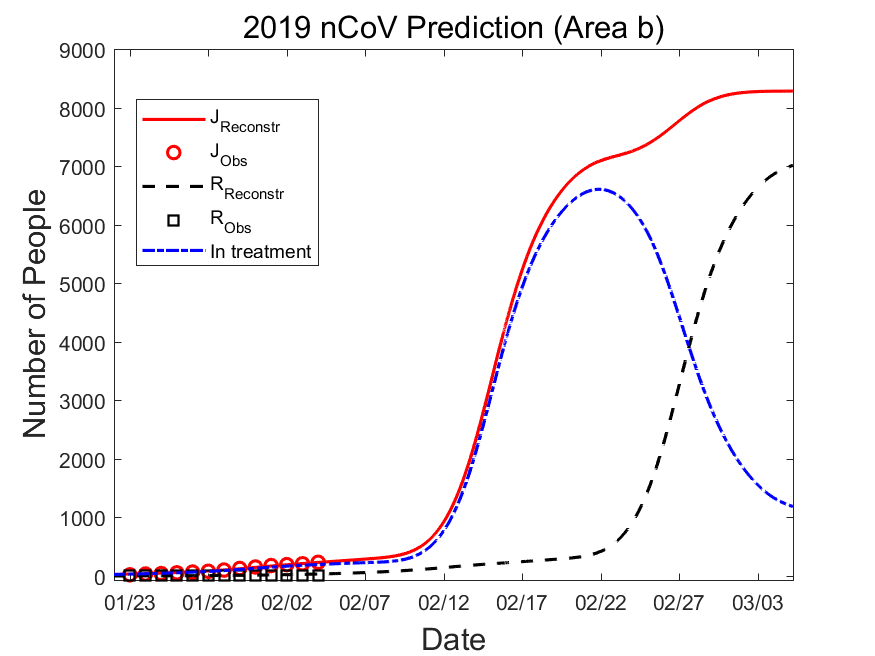}
			\label{ideal-34}
		}
	\end{center}
	\vskip -0.5 truecm
	\caption{\em Prediction with external source.}
	\label{Ideal5}
	\vskip 0.3 truecm
\end{figurehere}

In addition, the ending time of this  pneumonia would be absolutely postpone, we show Table \ref{tab::source} for reference. We shall remark that the same $\beta_b=0.2413$ and $\ell_b=0.5384$ are employed for the prediction.
\begin{tablehere}
	\begin{center}
		\begin{tabular}{ccc}
			\hline
			$\theta$ & cumulative confirmed people\\
			\hline
			$0\%$ &  $\approx$\;\;500 \\
			$1\%$ &   \;\;$\approx$\;\;1000 \\
			$5\%$  &  \;\;$\approx$\;\;4000 \\
			$10\%$  & \;\;$\approx$\;\;7500 \\
			\hline
		\end{tabular}
	\end{center}
	\vskip -0.4 truecm
	\caption{\em Prediction of cumulative confirmed people at the beginning of March.}
	\label{tab::source}
	\vskip 0.3 truecm
\end{tablehere}

Now the question comes to how to decrease the final cumulative confirmed people since the external source is inevitable. One feasible way is the increase of isolated ratio $\ell_b$, and the numerical simulations are shown in Figure \ref{Ideal7}. We can note that the cumulative confirmed people are sharply reduce with the high rate of isolation.
\begin{figurehere}
	\begin{center}
		\subfigure[$\ell_b=0.5384$]
		{
			\includegraphics[width=0.45\textwidth]{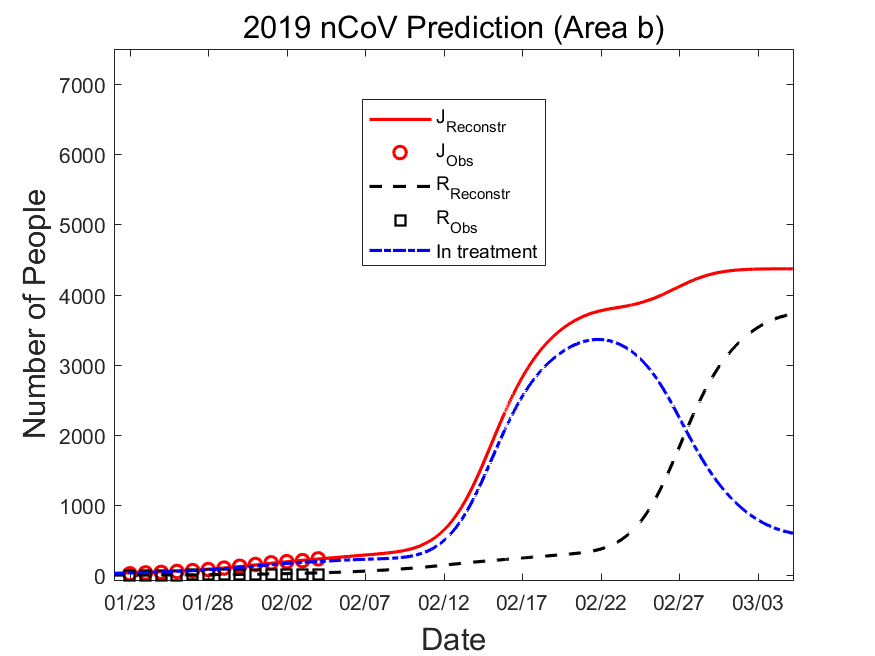}
			\label{ideal-71}
		}
		\subfigure[50\% $\ell_b$]
		{
			\includegraphics[width=0.45\textwidth]{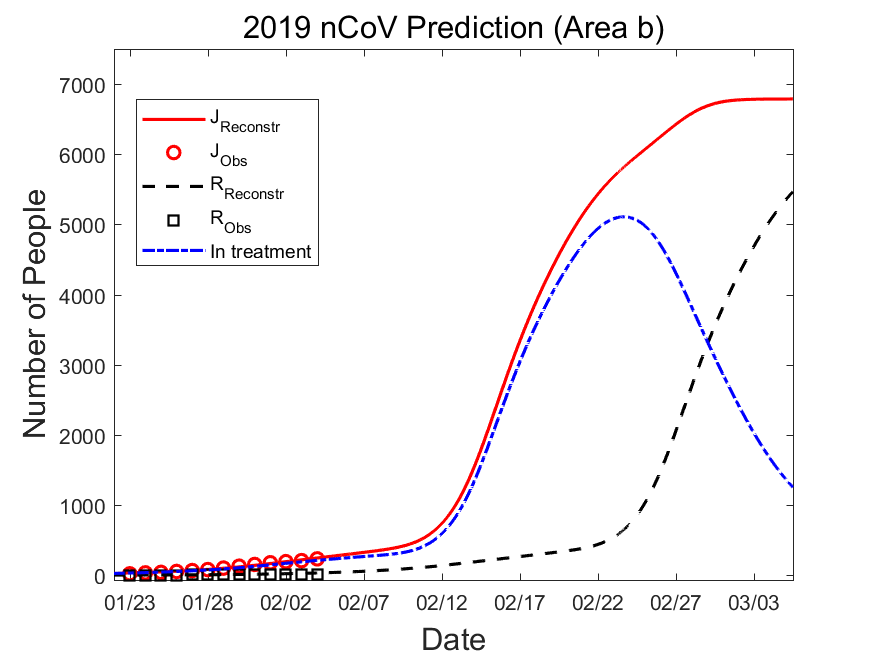}
			\label{ideal-72}
		}
		\\
		\subfigure[90\% $\ell_b$]
		{
			\includegraphics[width=0.45\textwidth]{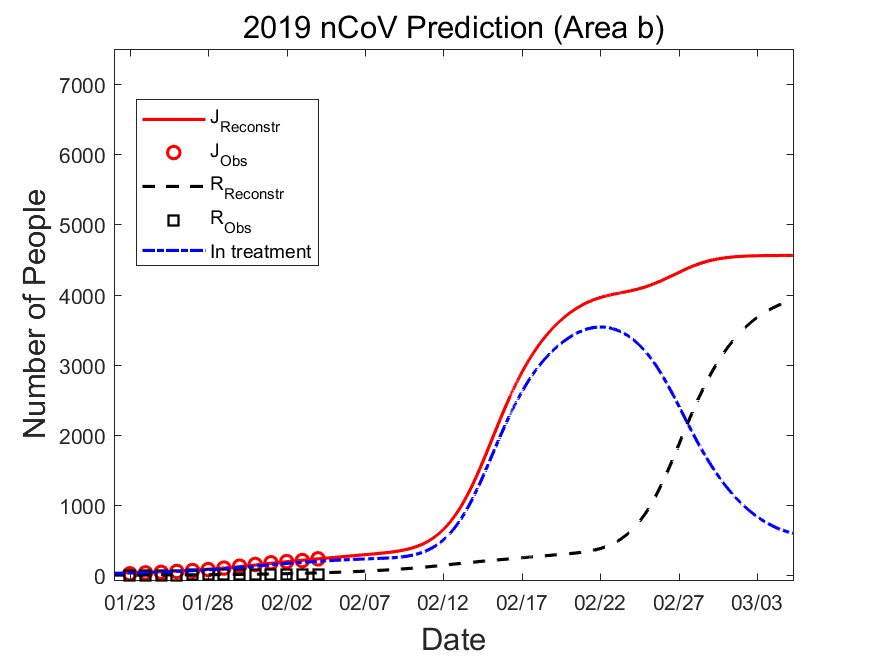}
			\label{ideal-73}
		}
		\subfigure[180\%$\ell_b$]
		{
			\includegraphics[width=0.45\textwidth]{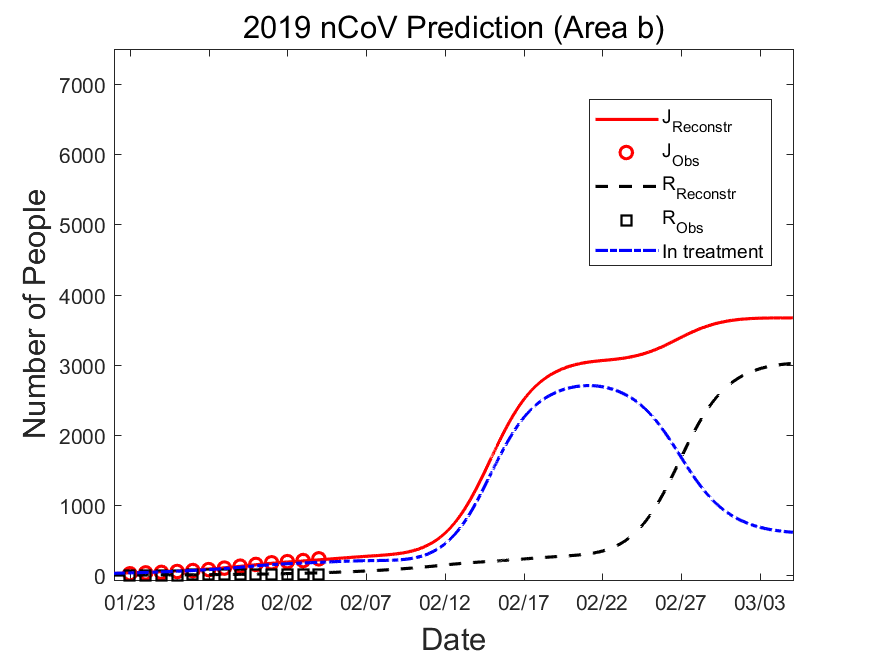}
			\label{ideal-74}
		}
	\end{center}
	\vskip -0.4 truecm
	\caption{\em Prediction with different impacts of isolated rate.}
	\label{Ideal7}
	\vskip 0.3 truecm
\end{figurehere}

In order to provide a better observation of the cumulative confirmed people with different impacts of isolated ratio, we exhibit the numerical results in Table \ref{tab::isolatione}.
\begin{tablehere}
	\begin{center}
		\begin{tabular}{ccc}
			\hline
			$\ell_b$ & cumulative confirmed people\\
			\hline
			$\ell_b$ &  $\approx$\;\;4500 \\
			50\% $\ell_b$&   $\approx$\;\;7000 \\
			90\% $\ell_b$  &   $\approx$\;\;4500 \\
	  	180\% $\ell_b$  &   $\approx$\;\;3500 \\
			\hline
		\end{tabular}
	\end{center}
	\vskip -0.3 truecm
	\caption{\em Prediction of cumulative confirmed people with different impacts of isolated rate.}
	\label{tab::isolatione}
	\vskip 0.3 truecm
\end{tablehere}

From the numerical simulations, we possess following conclusions
\begin{enumerate}
       \item The external source plays a significant role in the dynamic system, and the prediction of outbreak with the external source is more reliable than the one without it.
	\item The inflow of exposed people would increase the final cumulative confirmed people, and the ending time of this pneumonia would postpone.
	\item To avoid of rapid increasing of final cumulative confirmed people, the local government need to implement some more efficient restrictive policies to maintain the rate of isolation.
\end{enumerate}

\section{Conclusions}\label{sec:con}
In this paper, we have proposed a novel time delay dynamic system with external source. In this system, the suspected people of Area {\bf a} transfer to Area {\bf b} is concerned, and it is more reasonable and appropriate than the one in \cite{ChenArxiv2020, Yan2020} to describe the trend of local outbreak for the 2019-nCoV.
The numerical simulations are carried out to verify the effectiveness and accuracy of the novel time delay dynamic system with external source. Moreover, the newly proposed dynamic system can approximate the true data quite well in this event, and it could further forecast the trend of local event.
From the numerical simulations, we would like to advice that the local government apply some more efficient and strict measures to maintain the rate of isolation. Otherwise the local cumulative confirmed people of 2019-nCoV might be out of control.

At present, the parameters involved in the novel dynamic system are time-independent. In the future work, the change in the impact of isolation and spread rate may be assumed to depend on time so as to improve the agreement between real data and estimated solution. Besides, the complex network and stochastic process would be concerned in our dynamic system and the machine learning techniques would be applied to provide a better prediction of tendency for the outbreak of 2019-nCoV.

\section{Acknowledgements}
This work of Jin Cheng was supported in part by the National Science Foundation of China  (NSFC: No. 11971121), and
the work of Keji Liu was substantially supported by the Science and Technology Commission of Shanghai Municipality under the ``Shanghai Rising-Star Program'' No. 19QA1403400.
The authors thank for the helpful discussions with Prof. Guanghong Ding, Prof. Wenbin Chen and Prof. Shuai Lu in Fudan University, Prof. Xiang Xu in Zhejiang University,  Dr. Yue Yan and Dr. Boxi Xu in Shanghai University of Finance and Economics, and the collection of official data by Mrs. Jingyun Bian. Cheer up Wuhan!

\end{document}